\documentstyle[12pt]{article}
\def\PL{{\sl Phys. Lett.} }
\def\NP{{\sl Nucl. Phys.} }

\def\IJMP{{\sl Int. J. Mod. Phys.} }
\def\PR{{\sl Phys. Rev.} }
\begin {document}
%\large

\thispagestyle{empty}
{\hfill  Preprint JINR E2-96-219}\vspace{1.5cm} \\
\begin{center}
{\Large \bf
The canonical quantization of the kink -- model \\
beyond the static solution}
\vspace{0.5cm} \\
A. A. Kapustnikov\footnote{E-MAIL: KALEXAND@UNI.DP.UA}\vspace{0.5cm}\\
{\it Department of Physics, Dnepropetrovsk University, }\\
{\it 320625 Dnepropetrovsk, Ukraine}\\

\vspace{0.5cm}
A. Pashnev\footnote{BITNET: PASHNEV@THSUN1.JINRC.DUBNA.SU}\vspace{0.5cm}\\
{\it JINR--Laboratory of Theoretical Physics} \\
{\it Dubna, Head Post Office, P.O.Box 79, 101 000 Moscow, Russia}

\vspace{0.5cm}
A. Pichugin\footnote{E-MAIL: PEACH@UNI.DP.UA}\vspace{0.5cm}\\
{\it Department of Physics, Dnepropetrovsk University, }\\
{\it 320625 Dnepropetrovsk, Ukraine}

\vspace{1cm}
{\bf Abstract}
\end{center}
A new approach to the quantization of the relativistic
kink - model around the solitonic solution is developed on the ground of
the collective coordinates method. The corresponding effective action
is proved to be the action of the nonminimal $d=1+1$ point - particle with
curvature. It is shown that upon canonical quantization this action yields
the spectrum of kink - solution obtained firstly with the help of the
WKB - quantization.

\vspace{0.5cm}
\begin{center}
{\it Submitted to Phys. Rev. D}
\end{center}
\vfill
\setcounter{page}0

\setcounter{footnote}0
\newpage

\begin{center}
{\bf 1. Introduction}
\end{center}

It is well-known that the particles spectrum
in quantum field theories
(QFT) possessing topologically-nontrivial solution for the
corresponding equation of motion may be obtained either by the
 semiclassical WKB-method [1], or with the help
of effective actions [2].  The latter describes the low-
energy dynamics of a stable time-independent solutions (solitons, for
instance) plus  small quantum oscillations around it. It was established
earlier that these actions at the classical level are written down as
minimal $p$-branes actions, supplemented with  some additional
nonminimal terms [3].
Depending on curvatures of the relevant world-volumes of $p$-branes,
such nonminimal terms may become important when the quantum corrections
due to the field fluctuations in a neighborhood of static solutions
are not negligible.
Upon quantization these actions allow us to reveal the true
quantum content of their prototype field theories.

This approach amounts to a nonlinear change of the
space-time variables in the action of given QFT-model with spontaneously
broken relativistic symmetry [4], [5].
As compared with the old one, the new set of space - time variables
has another form of the transformations under the original space - time
symmetry group. In what follows the static solutions of the
QFT - model under
consideration become completely covariant with respect to the relativistic
group trasformations. To achieve this the special
dependence of the afore - mentioned nonlinear change of the space - time
variables on the
collective coordinates of solitons has been assumed.
In order to provide the canonical
relationship between these variables and the Goldstone excitations of
the spontaneously broken space - time symmetries
the original set of field variables
should be covariantly constrained.
Sometimes this is achieved by
putting all the fields with zero
expectation values equal to zero.
Thus we get the minimal $p$-brane action as the
residual effective action for the QFT-model we start with.
However, as it was mentioned above,  this procedure is not suitable for
models in which quantum perturbations of fields near the static
solutions are rather essential. In these cases one is compelled to be more
thorough in treating the excited modes of static solutions.

In this article we discuss the simplest example of  such a kind of
models -- the $d=1+1$ relativistic model $\phi^4$ -- in which the problem 
of exciting modes can be resolved both on the classical and quantum levels
simultaneously.

The outline of the paper is as follows. In Section 2 we review some well -
known results about the field - dependent transformations for the action of
the model under consideration.
It is shown how this transformation is associated with
the spontaneously broken relativistic symmetries of the initial action
and how the latter can be recast in the corresponding splitting form.

In Section 3 we have succeeded to show that the equation of motion obtained
from the "splitting" action for the perturbations of field
about kink - solution is covariantly separated into the equivalent system
of two equations. One of them is solved by the excited modes of the
static solution, while the other one defines a world - line trajectory of 
the
effective nonminimal point - particle in the weak curvature limit.

The implications of this solution for the construction of the corresponding
effective action from the original field action are proposed
in Section 4.

In Section 5 we extend this approach to
the quantum level. Here the canonical quantization of the effective action
is carried out for obtaining the quantum states of kink as the spectrum of
the underlying point - particle with curvature.

We finish with some speculations about the possible generalization of our
results to more complicated cases.

\begin{center}
{\bf 2. The splitting form action for kink}
\end{center}

To show more precisely how the collective coordinate method works,
let us consider the
action
\begin{equation}
S[\phi] = \int d^2 x \ L(\phi, \ \partial_m \phi)               \label{1}
\end{equation}
\begin{equation}
L(\phi, \ \partial_m \phi) = \frac{1}{2} (\partial_m \phi)^2 - 
\frac{1}{4}g^2 \Bigl[ \phi^2
- (\frac{m}{g})^2 \Bigr]^2,                                      \label{2}
\end{equation}
where $\phi(x,t)$ is a dimensionless scalar field; 
$m$ and $g$ are real parameters.
It is well - known that the corresponding equation of motion [6]
\begin{equation}
\partial^m \partial_m \phi + g^2 \Bigl[ \phi^2
- (\frac{m}{g})^2 \Bigr] \phi = 0,                            \label{3}
\end{equation}
leads to the kink solution
\begin{equation}
\phi_c (x) = \frac{m}{g} th \frac{mx}{\sqrt{2}},               \label{4}
\end{equation}
the time - independent solution,
which describes the bend of the field at the point
$x = 0$, with the width of order $m^{-1}$, and the nontrivial behaviour at
infinity
\begin{equation}
\phi_c( + \infty) = - \phi_c( - \infty) = \frac{m}{g}.          \label{5}
\end{equation}
Using the semiclassical method, it was found in [1], that
upon quantization the kink-solution (4) yields
a heavy quantum-mechanical
particle. The modified approach which we follow here shows that the same
result may be obtained straightforwardly from the
corresponding effective action. The way in which such a nonlinear
action would arise from the QFT-model (1),(2) can be seen by the following
considerations, analogous to those which were implemented
for the rigid string in
[4]. Let us pass to a new set of the basis variables in the action~(1)
\begin{eqnarray}
x^m &=& x^m(s) + e^m_{(1)}(s) \rho \equiv y^m(\sigma_A), \ \ \ A =\ 0, 
\ 1 \nonumber \\
\phi(x,t) &=& \widetilde{\phi}(\sigma_A),
\ \ \ \ \sigma_{A=0} = s, \ \ \ \ \sigma_{A=1} = \rho          \label{6}
\end{eqnarray}
where $x^m(s)$ are the coordinates of the $d = 2$ point-particle;
$e^m_{(1)}(s)$ is the unit space-like vector orthogonal to its world-line;
$\widetilde{\phi}(\sigma_A)$ is the notation for the field $\phi(x,t)$
in terms of the new variables. It is clear that it describes an infinite
set of the world-line fields emerging in its expansion with respect
to the inert coordinate $\rho$. It is worth to notice that unlike $(x,t)$
the new basis variables $(s,\rho)$ are invariant under Poincar\'{e}-
transformations. The reason of such behaviour of the 
variables is explained completely by the presence of
the coordinates of the point-particle $x^m(s)$ in the mapping (6). 
A specific feature of these coordinates is that the Poincar\'{e} -
translations
 symmetries associated
with the action (1)
\begin{eqnarray}
{x^m}' &=& x^m + \alpha^m,                                \nonumber \\
\phi'(x',t') &=& \phi(x,t),                     \label{7}
\end{eqnarray}
are realized on them in the form
\begin{eqnarray}
{x^m}'(s) &=& x^m(s) + \alpha^m,                            \nonumber \\
\widetilde{\phi}'(\sigma) &=& \widetilde{\phi}(\sigma),    \label{8}
\end{eqnarray}
indicating that the above symmetries are spontaneously broken with $x^m(s)$
thought of as the corresponding Goldstone fields
(owing to the nonhomogeneous part in
their transformation laws typical to the Goldstone fields). Besides,
one can check that due to the presence of the variables $e^m_{(1)}(s)$
in (6)
the Lorentz symmetry of the action (1) is spontaneously
broken  as well. So,
the only unbroken symmetry we are left with in the eqs.(6) is a gauge
symmetry under a world-line reparametrisation of the point-particle
\begin{eqnarray}
{x^m}'(s') &=& x^m(s),                                    \nonumber \\
s' &=& s'(s).                   \label{9}
\end{eqnarray}
The existence of this symmetry guarantees that after gauge fixing
\begin{eqnarray}
x^0(s) = t,\ \ \ \ \ x^1(s) = \widetilde{x}(t)              \label{10}
\end{eqnarray}
we get the set of physical degrees of freedom needed to recover
the field content of the original theory. To see this we make the change of
variables (6) in the action (1). This gives
\begin{equation}
S[x^m,\widetilde{\phi}] = \int d^2 \sigma \ \Delta(\sigma)
    \ L(\widetilde{\phi}, \ \nabla_m \widetilde{\phi} ),     \label{11}
\end{equation}
where we use the notations
\begin{eqnarray}
\Delta(\sigma) &=& det \frac{\partial y^m}{\partial \sigma^A} =
      \sqrt{\dot{x}^2} \biggl(1 - \rho k \biggr) \label{12} \\
L(\widetilde{\phi}, \ {\nabla}_m \widetilde{\phi} ) &=&
 \frac{1}{2} \biggl[ \frac{1}{\Delta^2} (\partial_{s} \widetilde{\phi})^2
\ -\ (\partial_{\rho} \widetilde{\phi})^2 \biggr] \ -\
\frac{1}{4} g^2 \biggl( \widetilde{\phi}^2 -
                          (\frac{m}{g})^2 \biggr)^2.             \label{13}
\end{eqnarray}
Note that the action (11) depends on the curvature of the
point-particle world-line
\begin{equation}
k = \sqrt{ - a^na_n}                       \label{14}
\end{equation}
where the acceleration $a^n$ is defined by
\begin{equation}
a_n = \frac{1}{\sqrt{\dot{x}^2}}\frac{d}{ds}\frac{\dot{x}_n}
{\sqrt{\dot{x}^2}}.    \label{15}
\end{equation}
It is rather evident that, in comparison to (1),
the action (11) involves one
superfluous degree of freedom $\widetilde{x}(t)$.
The latter is nothing but
the linear prototype of the Goldstone field $x(t)$, arising in the
decomposition of $\phi(x,t)$ around its static solution (4)
\begin{equation}
\phi(x,t) = \phi_c(x) - x(t){\phi}'_c(x) + ... ,           \label{16}
\end{equation}
To avoid the doubling of the
Goldstone degrees of freedom in the action (11) we are forced to relate
both $x(t)$ and $\widetilde{x}(t)$ fields by an equivalence transformation.
\newpage
\begin{center}
{\bf 3. Eliminating excited modes}
\end{center}

Here we wish to show that the most correct way to reveal this relation in
the framework of the model under consideration is to restrict the set of
variables $[x^m,\widetilde{\phi}]$ in the action (11) by the covariant
condition
\begin{equation}
\frac{\delta S[x^m,\widetilde{\phi}]}{\delta
(\delta \widetilde{\phi}(\sigma))}|_{x^m(s)=const} = 0   \label{17}
\end{equation}
where $\delta \widetilde{\phi}(\sigma)$ denote the perturbations of the
$\sigma$-fields $\widetilde{\phi}(\sigma)$ near the kink- solution
\begin{equation}
\widetilde{\phi}(\sigma) = \phi_c(\rho) + \delta \widetilde{\phi}(\sigma) 
  \label{18}
\end{equation}
It is worthwhile to emphasize that when varying the variable
$\delta \widetilde{\phi}(\sigma)$ in (17)
 one should treat both the $x^m(s)$ and
perturbations fields on equal footing as independent variables. Only then
can the equation (17) be considered as
the covariant constraint for eliminating
$\delta \widetilde{\phi}(\sigma)$ in terms of $x^m(s)$ and their
 derivatives. The
reason for this is rather simple and belongs to some common issues of the
effective actions theory [7]. We wish to briefly concern 
ourselves with this problem, with
emphasis on the peculiarities brought about by our model.

Let an action $S = S[x,X]$ which depends on two
sets of variables $\{x\}$ and $\{X\}$
achieve its extremum at the stationary point $x_c,X_c$. That is
\begin{equation}
\frac{\delta S}{\delta x}|_{x_c,X_c} = 0, \ \ 
  \frac{\delta S}{\delta X}|_{x_c,X_c} = 0.
\end{equation}
The corresponding effective action of the model is obtained by putting
\begin{equation}
S_{eff}[x] \equiv S[x,X(x)],
\end{equation}
were the constraint
\begin{equation}
\frac{\delta S}{\delta X}|_{x=const} = 0,\;\;  \Longrightarrow      
X = X(x)
\end{equation}
is used for excluding the subset of variables $\{X\}$ in terms of $\{x\}$.
Now any stationary point of the effective action
(20), with $X(x)$ subject to the constraint (21), is
simultaneously an
extremum of the total action $S[x,X]$. Indeed, varying the definition (20)
with respect to $x$, we find
\begin{equation}
\frac{\delta S_{eff}[x]}{\delta x} = \frac{\delta S[x,X(x)]}{\delta x} +
\frac{\delta S[x,X(x)]}{\delta X(x)}\frac{\delta X(x)}{\delta x}.
\end{equation}
The second term on the r.h.s. of eq. (22) disappear owing to (19), 
while the
first one vanishes only at the solution $x= x_c$
extremizing the effective action
\begin{equation}
\frac{\delta S_{eff}[x]}{\delta x} = 0.
\end{equation}
In our case the constraint (21) is represented by the manifestly covariant
condition (17). Let us analyze the structure of this condition in more 
detail.

By putting (18) into the action (11) one finds
\newpage
\begin{eqnarray}
S[x^m,\phi_c+\delta \widetilde{\phi}] & = & \int d^2 
\sigma \ \Biggl[\Delta \Biggl[ L(\phi_c) +
\frac{1}{2 \Delta^2} \Bigl( \partial_s \ \delta \widetilde{\phi} \Bigr)^2
   - \frac{1}{2} \Bigl( \partial_{\rho} \ \delta \widetilde{\phi} \Bigr)^2
 - \nonumber \\
& & - \frac{1}{2} g^2 \biggl( 3 \phi_c^2 -
 \Bigl( \frac{m}{g} \Bigr)^2 \biggr) \delta \widetilde{\phi}^2
    -  g^2 \phi_c \delta \widetilde{\phi}^3 \nonumber \\
& &  - \frac{1}{4} g^2  \delta \widetilde{\phi}^4 \Biggr]
+ {\phi}'_c \delta \widetilde{\phi} \partial_{\rho} \Delta \Biggr],
 \ \ \ \ \ \ \ \ \
{\phi}'_c \equiv \partial_{\rho} \phi_c,                \label{24}  \\
L(\phi_c) &=& - \frac{1}{2} {{\phi}'_c}^2
 - \frac{1}{4} g^2 \biggl( \phi_c^2 -
      \Bigl( \frac{m}{g} \Bigr)^2 \biggr)^2,                \label{25}
\end{eqnarray}
where we have used the equation of motion for $\phi_c$
\begin{eqnarray}
- {\phi}''_c + g^2 \phi_c \biggl( \phi_c^2 -
           \Bigl( \frac{m}{g} \Bigr)^2 \biggr) = 0.          \label{26}
\end{eqnarray}
From (24) it follows that the perturbations (18) are governed by the
equation
\begin{equation}
\Biggl[ \partial_s \Delta^{-1} \partial_s -
     \partial_{\rho} \Delta \partial_{\rho} +
g^2 \Delta \biggl( 3 \phi_c^2 - \Bigl(\frac{m}{g} \Bigr)^2 \biggr)
\Biggr] \delta \widetilde{\phi} + \phi_c' k\sqrt{\dot{x}^2} +
O(\delta \widetilde{\phi}^2) = 0.
\label{27}
\end{equation}
After the field rescaling $\delta\widetilde{\phi} \rightarrow
(m / \! g) \delta X$
and variable redefinition $\rho \rightarrow \epsilon u, \epsilon \equiv
\sqrt{2} / \! m $ equation (27) can be rewritten in a more convenient form
\begin{eqnarray}
& & \Biggl[\epsilon^2 \partial_{s} \Delta^{-1} \partial_{s} -
     \partial_u \Delta \partial_u +
  2 \Delta \biggl( 3X^2_c - 1 \biggr) \Biggr] \delta X +     \nonumber \\
& & \epsilon X_c' k\sqrt{\dot{x}^2} + O(\delta X^2) = 0,    \nonumber \\
& & \Delta =  \sqrt{\dot{x}^2} \biggl( 1 - \epsilon u k \biggr),
  \nonumber \\
& & X_c =  \tanh u,                \label{28}
\end{eqnarray}
where prime indicates the differentiation by $u$. Even in a linear
 approximation
when $O(\delta X^2) = 0$ this equation is rather complicated mainly due
to its
nonlinearities. So, we are not able to solve it unless some additional
assumptions are adopted. Firstly, we suppose that
the corresponding $\it{Ansatz}$ is given by
\begin{equation}
\delta X(s,u) = \epsilon k(s)f(u).        \label{29}
\end{equation}
Note that the zeroth-order solutions of eq.(29) are omitted owing to
nonhomogeneity term on the l.h.s., which is of first order in $\epsilon.$

Now we can  obtain approximate equations for $f(u)$ and $k(s)$ simply
expanding eq.(29) in Taylor series around $\epsilon = 0.$
In doing so we need, however, to be sufficiently careful in treating 
the first term inside the square brackets in (28). One cannot drop it
out as the terms of the second-order
correction in $\epsilon$ in the case when the point - particle
world-line metric is a quickly varying function.
Let us suppose that the term
\begin{equation}
\lim\limits_{\epsilon \to 0}
{\epsilon}^2 \frac{d}{ds}
\frac{1}{\sqrt{\dot{x}^2}} \frac{dk}{ds} \not= 0,
\end{equation}
is not zero even in the $\epsilon = 0$ limit. Taking into account
(29) and (30) one finds that to the lowest order in $\epsilon$ the
variables in eq.(28) are separated and  instead of one equation
with partial derivatives the equivalent system of two equations for total
derivatives arises
\begin{eqnarray}
{\epsilon}^2 \frac{1}{\sqrt{\dot{x}^2}} \frac{d}{ds}
\frac{1}{\sqrt{\dot{x}^2}} \frac{dk}{ds} +ck &=& 0,    \\
-f'' + 2(3X_c^2 -1)f - cf + {X'}_c &=& 0.
\end{eqnarray}
Here $c$ is a  constant which can be found as a solution of the
eq.(32). Both equations of the system
(31),(32) are very important for the constitution of the relationship 
between
linear and nonlinear parametrizations of the action. Firstly, it appears 
that
in contrast with the linear version of the theory the lowest eigenvalue of
the energy
is not zero. Indeed, one can see
that for $c = 0$  eq.(32) can be rewritten as
\begin{equation}
\frac{d}{du} \biggl[ \frac{1}{ \cosh^4u} \frac{d}{du}
\bigl( \cosh^2 u f_0 \bigr) \biggr] = \frac{1}{ \cosh^4u}.
\end{equation}
The corresponding solution can be represented in the form
\begin{eqnarray}
f_0(u) &=& \frac{1}{4} \cosh^2u - \frac{1}{12} \frac{\sinh^4u}{\cosh^2u} +
               \frac{b}{\cosh^2u}         \nonumber \\
& & + \frac{a}{8 \cosh^2u} \bigl( 3u + 2 \sinh 2u +
\frac{1}{4} \sinh 4u \bigr).                                    \label{34}
\end{eqnarray}
It is evident, that this solution ceases to have an appropriate
behavior for large $|u|$.
Thus one could not provide the stabilization of
the theory until this state is
assumed to be rejected from the very beginning. It is not hard to do this,
however, in the case $c \neq 0$. To this end, we have to go to the shifted
function
\begin{equation}
\tilde f(u) = f(u) - c^{-1} {X'}_c (u)                          \label{35}
\end{equation}
and carried out the calculation for $\tilde f(u)$ and $c$ from the
corresponding equation of motion
\begin{equation}
- \tilde {f''} +2 \bigl( 3X^2_c -1 \bigr) \tilde f - c \tilde f = 0.
 \label{36}
\end{equation}
Here we display the partial solution which is the most suitable for our
purposes
\begin{equation}
\tilde {f_1} (u) = \frac{\sinh u}{\cosh^2u}, \ \ \ c_1 = 3.    \label{37}
\end{equation}
It is appropriate to remark that besides (37) there exist other
solutions of the
eq.(36) [6]. However, we do not consider such solutions here
because they represent more excited eigenstates of the energy.

In the remainder of this section we want to discuss
the relationship between collective coordinates $x^m(s)$ and the Goldstone
field $x(t),$  which arises in the action upon spontaneous
breakdown of the
translation invariance. Let us note firstly that
we can take the decomposition (18) in the same form as
in eq. (16)
\begin{equation}
\tilde \phi(\rho ,t) = \phi_c(\rho) - g(s){\phi}'_c(\rho) + ... ,
 \label{37.1}
\end{equation}
where $g(s)$ as opposed to $x(t)$ is a completely covariant degree
of freedom. Comparing this expression with (6) in the proper - time gauge
(10) we find
\begin{equation}
g(t_r) = x(t_r) - \tilde x(t_r) + ...,  \label{37.2}
\end{equation}
where $t_r= t- \rho$ is the "retentive" time and dots stand for terms which
are at least second order in the field $\tilde x(t_r)$.
On the other hand from the
 solution (29), (35), (37) it follows that
$g(s) = - (2 g/3 m^3)\ k(s).$
Thus owing to the constraint
(17) there is no independent degree of freedom to be related with the zero
mode $\phi'_c(\rho)$. Instead of this we
have the relation
%\begin{equation}
$$\tilde x(t_r) = x(t_r) +
\frac{2}{3m^2} (\frac{g}{m})\ k(t_r) + ..., \nonumber \\
$$
%\end{equation}

%$g(t_r) \sim k(t_r)$
which establishes
the desired equivalence
connection between the Goldstone field $x(t)$ and corresponding collective
coordinate.

\begin{center}
{\bf
4. The effective action for kink with quantum corrections}
\end{center}

Thus we succeeded in resolving the constraint (17) by the Ansatz (29),
(30). Actually, this construction allows us to eliminate
$\delta \tilde \phi(\tilde{x})$ in terms of $x^m(s)$ and its derivatives.
As a by-product,
  the new equation (31) implementing the role of the low-energy
dynamical equation of motion for kink in presence of its excited modes
was obtained.
On the other hand one can consider the solution (29), (37) for
elimination of the
excited modes of the field $\tilde \phi(\tilde{x})$
from the action. Indeed,
putting together eqs.(29),
(31) and (37) and inserting them back into the action (24) we find
the following expression for the effective action
\begin{equation}
S_{eff} = - \mu \int ds \ \sqrt{\dot{x}^2} \ \biggl( 1
            + \frac{1}{3m^2} k^2 \biggr),                \label{38}
\end{equation}
where
\begin{equation}
\mu = \frac{2 \sqrt{2}}{3} \frac{m^3}{g^2}.                 \label{39}
\end{equation}
Note that in deriving  (\ref{38}) the $\rho$-integration in the action (24)
(modulo terms of order $k^3$) was
performed.

It was shown in Section 3 that the method at hand ensures the consistency 
of
a given action with the action (24). Namely, both of them possess
one and the same set of stationary points. To check the correctness of
this statement it is sufficient to compare eq.(31) with the equation of
motion derived from the effective action (\ref{38})
\begin{eqnarray}
\dot{p}_n & = & 0  \nonumber \\
p_n & = & \frac{\partial{L_{eff}}}{\partial{\dot{x}^n}} -
\frac{d}{ds}\frac{\partial{L_{eff}}}{\partial{\ddot{x}^n}}. \label{40a}
\end{eqnarray}
An important property of $L_{eff}$ defined from (\ref{38}) is that the 
eqs.(\ref{40a})
admit a very nice representation in terms of the world-line curvature 
$k(s)$.
The transition to such representation is achieved through the implication
 of
the corresponding Frenet equations [8]:
\begin{eqnarray}
\dot{e}^{m}_{(0)} & = & - k\sqrt{\dot{x}^2}e^{m}_{(1)},     \nonumber \\
\dot{e}^{m}_{(1)} & = & - k\sqrt{\dot{x}^2}e^{m}_{(0)},     \label{40}
\end{eqnarray}
where $e^{m}_{(a)}, a = 0,1$ are the basis components of the "moving" frame
\begin{eqnarray}
e^{m}_{(a)} & = & \biggl(e^{m}_{(0)} = \frac{\dot{x}^m}{\sqrt{\dot{x}^2}},
      \ \ e^{m}_{(1)} = - \frac{a^m}{k} \biggr),  \nonumber \\
e^{m}_{(a)}e_{m(b)} & = & \eta_{(ab)} = diag( + 1, - 1 ),    \nonumber \\
e^{m(a)}e^{n}_{(a)} & = & g^{mn} = diag( + 1, - 1 ).             \label{41}
\end{eqnarray}
Now it is not hard to show that making a projection 
of eq.(\ref{40a}) onto the
basis (\ref{41}) and taking account of eqs.(\ref{40}), we obtain
the equation of motion in the following compact form
\begin{equation}
\epsilon^2\frac{1}{\sqrt{\dot{x}^2}}\frac{d}{ds}
\frac{1}{\sqrt{\dot{x}^2}}\frac{dk}{ds}
+ \bigl( 3 - \frac{\epsilon^2}{2}k^2\bigr)k  =  0. \label{42}  \\
\end{equation}
Evidently, the eq.(31) is merely the linearized version of eq.(\ref{42}).

\begin{center}
{\bf
5. Quantization}
\end{center}

So far our attention has been devoted to the classical subject. 
It was shown
that the quantum field theory action (1), (2) around the topological defect
(4) can be reduced to the effective action (\ref{38}) which incorporates
implicitly the dependence of the theory on the zero and non - zero modes. 
The
latter are apparent themselves only 
from the mass formula (\ref{39}) which depends
inversely on the coupling constant $g.$ Thus we get, in fact, the new
non - perturbative formulation of the theory which may be regarded as
fundamental in its own right. In particular, it will be of great
interest  to use the action (\ref{38}) straightforwardly for obtaining the
quantum states of kink as the spectrum of point particle with
curvature.

We begin with some general comments about the canonical analysis of
this system. The Lagrangian (\ref{38}) depends, apart from the velocity
$\dot{x}_m,$ on the acceleration of the particle (15). This in accordance
with theories with higher derivatives means that we have to treat the
variables $x_m$ and $\dot{x}_m$ as canonically - independent
coordinates [9].
So, the phase space of our model consists of two pairs of the canonical
variables
\begin{eqnarray}
x_m, \ \ \ \ && p_m = \frac{\partial{L_{eff}}}{\partial{q^m}}
                          - \dot{\Pi}_m  ,    \label{44} \\
q_m = \dot{x}_m, && \Pi_m =
            \frac{\partial{L_{eff}}}{\partial{\dot{q}^m}}.
                                                        \label{45}
\end{eqnarray}
The explicit form of the momenta (\ref{44}) and (\ref{45}) are given by
\begin{eqnarray}
p^n &=& - e_{(0)}^n \bigl( \mu - \alpha k^2 \bigr) +
    \frac{2 \alpha e_{(1)}^n}{\sqrt{q^2}} \dot{k},  \label{46} \\
\Pi^n &=& - \frac{2 \alpha e_{(1)}^n}{\sqrt{q^2}} k,  \label{47}
\end{eqnarray}
where we use the notations
\begin{eqnarray}
k = \frac{\dot{q}_n e_{(1)}^n}{q^2}, \ \ \ \ \ \
                  \alpha = \frac{\mu}{3 m^2}. \label{48}
\end{eqnarray}
The components of the Frenet basis $e_{(a)}^n$ in the expressions 
(\ref{46}) --
(\ref{48}) are defined in (\ref{41}). This allows us to establish the 
existence of two
primary first - class constraints
\begin{eqnarray}
\Phi_1 &=& \Pi q \approx 0 \label{49} \\
\Phi_2 &=& H_{can} = pq + \Pi \dot{q} - L  \nonumber \\
 &=& pq + \sqrt{q^2} \bigl( \mu + \frac{1}{4 \alpha} q^2 \Pi^2 \bigr)
                   \approx 0 .       \label{50}
\end{eqnarray}
One verifies easily that upon substitution of
 the functions (\ref{46}), (\ref{47})
into (\ref{49}), (\ref{50}) these constraints vanish
identically with respect to
$q_n$ and $\dot{q}_n.$ This means that in fact the dynamics of our 
system is
constrained on a certain submanifold of the total phase space.
For definition of the physical phase space according to Dirac's
gauge fixing prescription [10]  we need to
introduce new constraints which enable us
to avoid the
gauge freedom of the theory generated by the constraints
(\ref{49}) and (\ref{50}).
To this end we shall consider only the ordinary proper - time 
gauge condition
\begin{eqnarray}
X = \sqrt{q^2} - 1 \approx 0. \label{51}
\end{eqnarray}
It is easy to see that $X$ forms
 a second - class
algebra with the constraint (\ref{49})
\begin{eqnarray}
\{ \Phi_1, X \} = 1 + X. \label{52}
\end{eqnarray}
Therefore they must be omitted as non - dynamical degrees of freedom.
This result can be achieved if we pass from the Poisson's brackets to Dirac's
ones
\begin{eqnarray}
\{ f, g \}^{\ast} &=& \{ f,g\} + \frac{1}{1+X} \{f, \Phi_1 \} \{X,g\}
 - \frac{1}{1+X} \{f,X \} \{\Phi_1, g \}. \label{53}
\end{eqnarray}
From (\ref{53}) one can observe that the 
constraints (\ref{49}), (\ref{51}) have zero Dirac's
brackets with everything and therefore can be considered as strong 
equations.

The remaining phase -- space coordinates have a transparent physical 
meaning.
Indeed, let us perform the transformation to the new set of canonically
conjugated variables
\begin{eqnarray}
\rho = \sqrt{q^2}, \ \ \Pi_{\rho} = \Pi_{(0)},  \nonumber \\
v = arcth \frac{p_{(1)}}{p_{(0)}}, \ \ \Pi_v = \rho \Pi_{(1)},\nonumber \\
z^m = x^m - \{x^m, v\} \Pi_v, \label{54}
\end{eqnarray}
where $p_{(a)} = p_n e_{(a)}^n,$ $\Pi_{(a)} = \Pi_n e_{(a)}^n$ are the
momentum components in the Frenet basis. We would like to note that 
the above
representation of the phase -- space variables preserves the structure of 
the
canonical Poisson brackets:
\begin{eqnarray}
\{\Pi_{\rho}, \rho \} &=& \{\Pi_v, v\} = 1 \nonumber \\
\{p^m, z_n \} &=& \delta^m_n  \label{55}
\end{eqnarray}
with all others equal to zero. It follows from their definitions that the
constraints (\ref{49}), (\ref{50}), (\ref{51}) when rewritten
in terms of new coordinates have
the form
\begin{eqnarray}
\Phi_1 &=& \rho \Pi_{\rho} \nonumber \\
\Phi_2 &=& \rho \bigl( -\sqrt{p^2}\ ch\ v + \mu
      - \frac{1}{4 \alpha} {\Pi_v}^2
       + \frac{1}{4 \alpha} \rho^2 \Pi_{\rho}^2 \bigr), \label{56} \\
X &=& \rho - 1. \nonumber
\end{eqnarray}
Thus we see that the physical variables $\Pi_v,$ $v$ are nothing but
the intrinsic angular momentum, i.e. the spin
\begin{eqnarray}
S_{mn} = q_m \Pi_n - q_n \Pi_m = \varepsilon_{mn} \Pi_v,      \label{57}
\end{eqnarray}
and the corresponding angular variable in this model. The residual
first - class constraint $\Phi_2$ in terms of the remaining coordinates
has the form
\begin{eqnarray}
\Phi_2 = - \sqrt{p^2}\ ch \ v + \mu - \frac{1}{4 \alpha} \Pi_v^2. 
\label{58}
\end{eqnarray}
The physical states on the quantum level must fulfill
the condition $\widehat{\Phi_2} |ph> = 0.$ 
Here the operator $\widehat{\Phi_2}$
is obtained from (\ref{58}) by the normal ordering prescription
in the coordinate representation $\Pi_v = -i\ \partial/ \partial v :$
\begin{eqnarray}
\widehat{\Phi_2} &=& a^{+}_{\lambda} a_{\lambda}
           - \xi^2 (1 - \lambda) \nonumber \\
a_{\lambda} &=& \frac{\partial}{\partial v}
   + \sqrt{2 \lambda} \xi sh \frac{v}{2},                 \label{59a} \\
\xi &=& \frac{2}{\sqrt{3}} \frac{\mu}{m},            \nonumber \\
\lambda &=& \frac{\sqrt{p^2}}{\mu} \equiv \frac{M}{\mu} \geq 0. \nonumber
\end{eqnarray}
Note, that there are other similar modifications of the operator
$\widehat{\Phi_2}$ admissible from the classical point of view. Depending
on the operators ordering procedure for the product of operators
$a_{\lambda},$ $a^+_{\lambda}$ they are not equivalent to the expression
(\ref{59a}).  We do not find them very satisfactory since in what follows
such representations lead to difficulties with 
the physical interpretation of the
system. In particular, the obvious interpretation of the
groundstate of kink as a state with $M=\mu$ does not hold.

Thus here we shall confine ourselves to the
equation of motion
\begin{eqnarray}
[ a^{+}_{\lambda} a_{\lambda} - \xi^2 (1 - \lambda) ] \Psi (v) = 0, 
\label{59b}
\end{eqnarray}
where the operators $a_{\lambda},$ $a^+_{\lambda}$ are defined in
(\ref{59a}).
This equation can be rewritten in the standard form of the Schr\"odinger
equation
\begin{eqnarray}
\biggl( - \frac{\partial}{\partial v^2} + 2 \lambda \xi^2 \ sh^2
\frac{v}{2} - \sqrt{\frac{\lambda}{2}}\ \xi \ ch \frac{v}{2} \biggr) \
\Psi (v) = \xi^2 (1 - \lambda) \Psi (v),                     \label{59c}
\end{eqnarray}
with the potential
\begin{eqnarray}
V(v) = 2 \lambda \xi^2 \ sh^2 \frac{v}{2} -
\sqrt{\frac{\lambda}{2}}\ \xi \ ch \frac{v}{2}                 \label{59d}
\end{eqnarray}
and the energy $E=\xi^2(1 -\lambda).$
The interesting feature of the potential $V(v)$ is that it can be
written in the form
\begin{equation}
V(v)=G^2(v)-G'(v), \; \; \ G(v)=\sqrt{2\lambda}\xi \ ch \frac{v}{2},
\end{equation}
typical to supersymmetric quantum mechanics [14].

One may check that for all the solutions
of eq. (\ref{59c}) with $\mu \geq m, \ \ \ (\xi \geq 1)$ the potential
(\ref{59d}) has only one stationary point $V_{min} = - \sqrt{\lambda/2}\
\xi.$ Hence, from the boundary condition $E \geq V_{min}$ we get the
following constraint on the permissible values of the parameters $\lambda$
and $\xi:$
\begin{eqnarray}
\xi (1 - \lambda)\ \geq - \sqrt{\lambda/2}.                 \label{59i}
\end{eqnarray}
We will come back to this point later on.

Let us now consider the eigenfunction which can be obtained from the eq.
(\ref{59c}). Firstly, it is easy to notice that there exists a 
well - defined
ground - state solution with the canonical eigenvalue $M=\mu \ \ (E=0):$
\begin{eqnarray}
a_{\lambda = 1}
\Psi_{vac} (v) = 0\ \Longrightarrow \ \Psi_{vac} = C\ exp \biggl( -2
 \sqrt{2}\ \xi\ ch \frac{v}{2} \biggr).                 \label{60}
\end{eqnarray}
In
order to construct the wave functions for other physical states it will be
convenient to perform the following standard decomposition
\begin{eqnarray}
\Psi (v) = W(v) \  exp \biggl( -2 \sqrt{2 \lambda} \ \xi \ ch \frac{v}{2}
\biggr).                                          \label{61}
\end{eqnarray}
When substituting (\ref{61}) into (\ref{59c}) a new equation for the
function $W(v)$  appears
\begin{eqnarray} W'' - 2 \sqrt{2 \lambda} \ \xi
\ sh \frac{v}{2}\ W' + \xi^2 (1 - \lambda) \ W = 0.  \label{eq.for.W}
\end{eqnarray}
There are only two forms of representation
for the function $W(v)$ compatible with the structure of eq.
(\ref{eq.for.W})
\begin{eqnarray}
W^{(+)} &=& \sum\limits_{n=0}^{\infty}\
A_n^{(+)} \ ch  \frac{nv}{2},                         \label{68}      \\
W^{(-)} &=& \sum\limits_{n=0}^{\infty}\
A_n^{(-)} \ sh  \frac{(n+1)v}{2}.                      \label{sum.for.W}
\end{eqnarray}
The first representation is even and the second one is odd 
under the transformation $v \rightarrow -v.$ For
definiteness let us consider the function $W^{(-)} (v).$ As a
result of the substitution of 
the decomposition (\ref{sum.for.W}) into the eq.
(\ref{eq.for.W}) we obtain the recursive conditions for the coefficients
$A^{(-)}_n :$
\begin{eqnarray} \sum\limits_{n=0}^{\infty} \ a_{mn}^{(-)}
A_n^{(-)} =0,                                    \label{recurA}
\end{eqnarray}
with
\begin{eqnarray}
a_{mn}^{(-)} =&-& \xi \bigl(
\lambda /2 \bigr)^{\frac{1}{2}} m\ \delta_{n,m-1} + [ \xi^2 (1 -
\lambda) +  (m+1)^2/4] \delta_{n,m} \nonumber \\
   &+& \xi \bigl(
   \lambda /2 \bigr)^{\frac{1}{2}} (m+2) \delta_{n,m+1}.
                                                       \label{coef.a.mn}
\end{eqnarray}
It is obvious, that in accordance with the
homogeneity of the system (\ref{recurA}) it appears the
equation
\begin{eqnarray}
det \  a_{mn}^{(-)} = 0,        \label{det.a}
\end{eqnarray}
to be fulfilled. It gives the relationship
between parameters $\lambda$ and $\xi.$ The roots of this
equation are exhibited in Fig. 1, where the eigenvalues of $\lambda^{(-)}$
are expressed as a function of $\xi.$
It is evident that only the part of the
dots, indicated by the solid line, has been interpreted as the physical
ones. The remaining branch, denoted by the dashed line, is not compatible 
with
the condition (\ref{59i}). Therefore the dot $\xi=\xi_c \cong 15 $ can be
regarded as the crucial one. In order to understand better this result it
would be useful to remind our previous lesson. It was mentioned in 
Section 3
that in accordance with [1] the weak - coupling
corrections to the kink solution can be summarized in the mass formula
\begin{eqnarray}
M = \mu + \frac{1}{2} \sqrt{\frac{3}{2}} \ m.        \label{mass.for}
\end{eqnarray}
In our notations ( see (\ref{59a}) ) this expression is given by
\begin{eqnarray}
\lambda^{(-)} = 1 + \frac{1}{\sqrt{2} \xi}.        \label{def.lam.xi}
\end{eqnarray}
It is of main importance that the last formula is
in a good agreement with all the numerical
data represented in Fig. 1 by the solid
line. From foregoing it follows that the point $\xi_c$ may be regarded as
thate in which the weak - coupling perturbative regime of the theory
is replaced by its strong - coupling regime and we left with
the solitonic solution only.

Having the spectrum of kink given by the explicit function
(\ref{def.lam.xi}) we can construct eigenfunctions
$\Psi^{(-)}$ straightforwardly by making use of the recursive condition
(\ref{coef.a.mn}). The results of this calculations for
$n$ ranging from 1 to 20 are drawn in Fig. 2. It is worthwhile to remark
that the function $W^{(-)} (v)$ we are dealing with is, in principle, a 
well
defined and calculable quantity, since the reduction of the
series (\ref{sum.for.W})
is provided by the recursive  condition (\ref{coef.a.mn}). It turns out
that for large $m$ and the function $\lambda^{(-)} (\xi)$ defined by
(\ref{def.lam.xi}) there  exists the simple expression for
the coefficients $A^{(-)}_n$ in eq. (\ref{sum.for.W}):
\begin{eqnarray}
\frac{A^{(-)}_m}{A^{(-)}_{m-1}} &=&
 \frac{\alpha_m}{1 + \alpha^2_m /1 + \cdots},  \nonumber \\
\alpha_m &=& \frac{2 \sqrt{2} \ \xi}{m}.               \label{Am.Am-1}
\end{eqnarray}
From (\ref{Am.Am-1}) it follows that the calculations of $W^{(-)} (v)$ are
consistently defined.

The above considerations are connected to the method of calculation of the
odd wave function $\Psi^{(-)} (v)$ only. In principle, the analogous method
allows to find a formal solution for the even wave function
$\Psi^{(+)} (v)$ also. The latter,
however, is not relevant from the physical point of view since the
corresponding eigenvalues equation $det \ a^{(+)}_{mn} = 0$ for the 
function
$\lambda^{(+)} (\xi)$ is in contradiction with the boundary condition
(\ref{59i}).

\begin{center}
{\bf
6. Conclusion}
\end{center}

A few comments are in order. In this paper we have constructed a new
nonperturbative approach to the problem of quantization of the
topologically -- nontrivial QFT -- models with spontaneously broken
relativistic symmetry. Ultimately this approach is based on the
observation that the spectrum of the localized field states can be restored
from the corresponding effective actions. The method of deriving the
effective actions from the field theory is proved to be
reduced to the transformation (6)
allowing to eliminate some field variables of the theory in terms of the
appropriate collective coordinates. It was shown that in the case of kink
solution this yields the action of the nonminimal $d=1+1$ point - particle
with curvature. Upon quantization it describes the quantum particle with
the mass (\ref{mass.for}) obtained earlier in the framework of the
WKB - approach [1].

It is clear, that the methods developed here can be extended to the cases 
of
field models involving more complicated static solutions. In particular,
it would be of great interest to apply this approach to 
carry out the quantization
of the t'Hooft - Polyakov monopole [11] and
of the static solution of (2+1) - dimensional $SU(2)$ Yang - Mills theory
with the Chern - Simons term [12]. We hope also that such an investigation
turns out
to be useful in the cases of the solutions obtained from the supersymmetric
field theories (see for example [13] and references therein).

This work has been supported in part by International Soros Science 
Education
Program of the International Renaissance Foundation
through grant N APU062045, the Russian Foundation of
Fundamental Research, grant 96-02-17634  and INTAS, grant 94-2317 and
grant of the Dutch NWO organization.

\newpage
\setlength{\unitlength}{1mm}

\begin{picture}(170,110)(0,130)
\put(15,230){\special{em:graph figure1.msp}}
\put(70,132.){Figure 1. }
\end{picture}
%\normalsize
\small
%\footnotesize

\begin{tabular}{|c||c|c|c|c|c|c|c|c|c|c|c|c|c|c|c|c|c|c|c|c|c|c|c|c|}
  \hline
$\xi$           & 3.16 & 3.87 & 4.47& 5.00& 5.48 & 5.92& 6.32& 6.71&
 7.07& 10.0& 12.0& 13.0 \\ \hline
$\lambda (\xi)$ & 1.55 & 1.425&1.36 & 1.32& 1.289&1.265&1.245& 1.23&
1.216& 1.15&1.144&1.115  \\ \hline \hline
$\xi$           & 14.0& 15.0& 16.0& 17.0& 18.0& 19.0& 20.0& 21.0&
 22.0& 23.0& 24.0& 25.0  \\ \hline
$\lambda (\xi)$ &1.09& 1.047&1.044&1.042&1.040&1.037&1.035&
1.033&1.032&1.030&1.029&1.028 \\ \hline
\end{tabular}

\large
~~~~~~~~~~~~~~~The eigenvalues of $\lambda^{(-)}$ as 
a function of the parameter $\xi.$ \\

\begin{picture}(170,90)(0,120)
\put(15,205){\special{em:graph figure2.msp}}
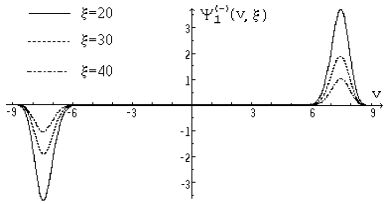
\put(10,125){Figure 2. The odd wave function $\Psi^{(-)}_1 (v, \xi)$ for
the excited kink  }
\put(10,118){state with $\xi =20,\ 30, \ 40.$}
\end{picture}

\end {document}